\documentclass[%
 preprint,superscriptaddress,
 amsmath,amssymb,
 aps,
]{revtex4-1}

\usepackage{graphicx}% Include figure files
\usepackage{dcolumn}% Align table columns on decimal point
\usepackage{bm}% bold math
\usepackage{threeparttable}
\usepackage{lipsum}
\usepackage{setspace}
\begin{document}

\title{Measuring national capability over big science's multidisciplinarity: A case study of nuclear fusion research}% Force line breaks with \\

\author{Hyunuk Kim}
\affiliation{Department of Industrial and Management Engineering, Pohang University of Science and Technology, Pohang 37673, Republic of Korea}
\affiliation{Kellogg School of Management, Northwestern University, Evanston, IL 60208, United States of America}
\affiliation{Northwestern Institute on Complex Systems, Evanston, IL 60208, United States of America}
\author{Inho Hong}
\affiliation{Department of Physics, Pohang University of Science and Technology,\\ Pohang 37673, Republic of Korea}
\author{Woo-Sung Jung}%
 \email{wsjung@postech.ac.kr}
\affiliation{Department of Industrial and Management Engineering, Pohang University of Science and Technology, Pohang 37673, Republic of Korea}
\affiliation{Department of Physics, Pohang University of Science and Technology,\\ Pohang 37673, Republic of Korea}
\affiliation{Asia Pacific Center for Theoretical Physics, Pohang 37673, Republic of Korea}

             
\begin{abstract}
In the era of big science, countries allocate big research and development budgets to large scientific facilities that boost collaboration and research capability. A nuclear fusion device called the ``tokamak'' is a source of great interest for many countries because it ideally generates sustainable energy expected to solve the energy crisis in the future. Here, to explore the scientific effects of tokamaks, we map a country's research capability in nuclear fusion research with normalized revealed comparative advantage on five topical clusters -- material, plasma, device, diagnostics, and simulation -- detected through a dynamic topic model. Our approach captures not only the growth of China, India, and the Republic of Korea but also the decline of Canada, Japan, Sweden, and the Netherlands. Time points of their rise and fall are related to tokamak operation, highlighting the importance of large facilities in big science. The gravity model points out that two countries collaborate less in device, diagnostics, and plasma research if they have comparative advantages in different topics. This relation is a unique feature of nuclear fusion compared to other science fields. Our results can be used and extended when building national policies for big science.
\end{abstract}

\maketitle

%\tableofcontents

\section{Introduction}
Big science is characterized by its big budgets, manpower, and machines. 
It includes a number of multidisciplinary fields such as nuclear fusion, particle accelerators, and space science~\cite{weinberg1961impact}.
Most of them originated for military reasons in World War II and were mainly led by superpowers. 
In recent decades, as these fields become more demanding, countries actively collaborate to utilize the resources of others and build shared infrastructure~\cite{capshew1992big,Xin1548,fortin2013big}.
In this sense, compared to little science, big science requires more international collaboration and resource accessibility~\cite{sonnenwald2007scientific}.

%limitation of current method, why is our study important?
A large facility is considered the core resource of big science. 
From construction to operation, it requires participation of various stakeholders under the leadership of national government, 
resulting in economic spillovers to society~\cite{autio2004framework,choi2017economic,castelnovo2018economic}. 
A large facility also stimulates scientific advancements by supporting research activities that are hard to conduct in a laboratory.
It attracts researchers of diverse disciplines and enhances scientific collaborations. 
Despite its scientific importance, little attention has been paid to examining how large facilities raise national research capacities because of difficulties in unraveling the multidisciplinarity of big science~\cite{heidler2015qualifying,hallonsten2016use,qiao2016scientific}. 
Moreover, national research capacity is difficult to quantify as it is built on the complex interactions between private and public domains~\cite{freeman1995national,etzkowitz2000dynamics}. 
Depending on science and technology policies, countries have different goals, such as training experts, publishing papers, or granting patents, that constitute the national research capacity~\cite{feller1997federal,laredo2001research}.

%\IH{Can we find some references on the last sentence?}
%high complexity  big science arising from intensive collaborations and multidisciplinarity~\cite{heidler2015qualifying,hallonsten2016use,qiao2016scientific}.}

%\IH{We need some reasoning on why we focused on academic publishing; the answer is quantification. For example, ``Among many aspects of national research capacity, we focus on academic publishing that can describe the capacity quantitatively. Here, we measure this national research capability by ... ''. The other sentences are good.}

Among many aspects of the national research capacity, this study focuses on academic publishing to estimate the capacity quantitatively \cite{doi:10.1177/0164025987009002006,borner2012design,sinha2015overview,wang2016large,chen2015identifying,guevara2016research,li2017evolutionary}, which we term ``research capability,'' by implementing topic modeling and revealed comparative advantage on the bibliographic information of research papers. The dynamic topic model~\cite{blei2006dynamic,gerrish2010language} first detects subject fields from paper abstracts and distributes publication counts over the detected fields in real values. Normalized revealed comparative advantage (NRCA)~\cite{yu2009normalized} is applied to fractional publication counts for projecting a country's research capability as well as its changes by facility construction. Based on NRCA, we measure how similar two countries' research capabilities are and include the distance in a gravity model to show its impact on international collaboration.

\clearpage

For a case study, we investigate nuclear fusion, in which the construction of large facilities and international collaborations are crucial. Nuclear fusion is a field that countries have interest in as it produces clean, affordable, and sustainable energy~\cite{chen2011indispensable,clery2013piece}. The history of nuclear fusion consists of the footprints of major successes in tokamaks~\cite{braams2002nuclear}.
After the nuclear fusion reaction of hydrogen was identified as the source of solar energy in the 1920s~\cite{nla.cat-vn2495749}, scientists began to study controlled thermonuclear fusion for sustainable energy production in the 1950s~\cite{smirnov2009tokamak}.
The tokamak is a device that magnetically confines high-temperature plasmas essential for steady thermonuclear reactions~\cite{wesson2011tokamaks}, and now it is the most dominant and actively studied device for nuclear fusion research~\cite{kikuchi2010review}.
Tokamaks are composed of strong magnets for confining plasmas, several wall-components in a vacuum vessel for protection, heating devices, and diagnostic devices, which require knowledge across diverse fields: plasma physics, numerical simulations, diagnostics, material science, and engineering~\cite{wesson2011tokamaks}.
The performance of tokamaks positively scales with size, thus tokamaks have become greater, better, and more expensive~\cite{lawson1957some,aymar2002iter,ikeda2009iter,grandoni2015}.
The large budgets for tokamaks have increased international collaborations since the 1990s, as seen in the cases of JET (Joint European Torus)~\cite{rebut1985joint} and ITER (International Thermonuclear Experimental Reactor) construction~\cite{aymar2002iter}. 

%summary of this paper
Our approach successfully captures various aspects of nuclear fusion from a bibliographic database over 40 years, 1976--2016. The dynamic topic model disentangles multidisciplinarity and classifies 41 topics grouped into five topical clusters: material, plasma, device, diagnostics, and simulation. Furthermore, the revealed comparative advantage identifies leading countries that participate in international projects or have their own tokamak. The rise and fall of these countries match well with tokamak operation. With the gravity model of scientific collaboration, we additionally address whether complementarity leads to collaboration in nuclear fusion research. The regression results show that countries collaborate less if they have research capability in different topics. It is a unique characteristic of nuclear fusion compared to other sciences in which complementarity enhances collaborations~\cite{oh2005coauthorship,barjak2008international,heinze2008across,acosta2011factors,zhang2017china}. This paper provides quantitative evidence for establishing strategic policies that initiate and evaluate big science projects~\cite{gerow2018measuring}. 
\clearpage
% novelty of the research
%% related conventional approach: metric, performance (?)
%%% knowledge structure: bibliometric studies -> in big sciences

%% issue-raising
%%% (one-sentence research question!)
%%% need to decompose the context into two dimensions: multidiscplinarity & nationality
%%% interdisciplinary aspect -> need of conceptualization of knowledge
%%% international collaboration -> need to measure comparative research performance of nations (RCA) -> no direct mentioning on RCA

%% new thing in this article: 
%%% context: thermonuclear fusion, tokamak
%%% methodology: conceptualization by topic modeling, ranking by RCA

%% expected result
%%% 1. concepts in the field are categorized, and grouped by their characteristics, such as plasma physics, engineering, material sciences, design and operations.
%%% 2. installation of tokamaks -> explains research power of nations.
%%% 3. (collaborative aspect)

% what we did in this paper: structure
%% brief introduction to data
%% topic modeling -> clustering into groups
%% RCA -> comparative research power across nations and research topics.
%% trend of ranks: rise and fall of performance by tokamaks.

%\section*{Literature Review}
%\subsection*{Big science, why is it important?}
%\subsection*{Big science project evaluation}
%\subsection*{Scientometric approaches}

% You may title this section "Methods" or "Models". 
% "Models" is not a valid title for PLoS ONE authors. However, PLoS ONE
% authors may use "Analysis" 
\section{Data and Methods}

\subsection{Bibliographic data}

We analyzed 25,085 nuclear fusion research papers published during 1976-2016. They were collected from the Scopus database (document type: article) and contain the term ``tokamak'' in the title, abstract, or keyword fields. Papers without affiliation information were manually filled by checking their original documents. When an author had multiple affiliations, we considered the first one as her/his nationality. We used the fractional counting method to obtain the number of papers for each country. For example, if a paper was written by three American and two Korean researchers, 0.6 and 0.4 were assigned to both countries' paper counts. 

The fractional counting method gives more weight to leading countries, so that would embrace their inherent academic leadership. Nevertheless, the fractional counting method gives less biased results than the full counting method that assigns an equal weight to all countries in a paper. The full counting method could overrepresent some countries (e.g. the United States) which participate in many international projects. Systemic comparisons of the two methods recommend the fractional counting method in co-authorship analysis~\cite{perianes2016constructing, park2016normalization}, especially for scientific fields conducting large-scale international experiments. For this reason, we chose the fractional counting method to estimate research capability as well as the degree of collaborations.

Among 75 countries in our dataset, we focused on the top 14 countries that published more than 250 papers in our time scope. The distribution of paper counts was highly skewed. These 14 countries published more than 90\% of the research articles. The top 14 countries were the United States, Japan, China, Germany, the United Kingdom, Russia, France, Italy, the Republic of Korea, Switzerland, India, Sweden, Canada, and the Netherlands. The basic statistics of these countries are listed in Table \ref{tab:country_pub}. A paper written by more than two authors in different countries is classified as a collaborative paper.

\begin{table}[!ht] 
\caption{Summary statistics of 14 leading countries in nuclear fusion research. All values are real numbers as we count the number of papers by the fractional counting method. Ratio is the proportion of collaborative papers to total papers.\\}
\footnotesize
\begin{tabular}{|l|l|l|l|l|}
\hline
\textbf{Country} & \textbf{Collaborative Papers} & \textbf{Total Papers} & \textbf{Ratio} \\ \hline
United States& 978.4& 7646.4& 0.13\\ \hline
Japan& 411.7& 3025.7& 0.14\\ \hline
China& 335.7& 2777.7& 0.12\\ \hline
Germany& 738.1& 2147.1& 0.34\\ \hline
United Kingdom& 522.5& 1775.5& 0.29\\ \hline
Russia& 299.5& 1392.5& 0.22\\ \hline
France& 403.6& 1135.6& 0.36\\ \hline
Italy& 325.1& 964.1& 0.34\\ \hline
Republic of Korea& 115.8& 424.8& 0.27\\ \hline
Switzerland& 153.5& 409.5& 0.37\\ \hline
India& 49.8& 400.8& 0.12\\ \hline
Sweden& 135.0& 326.0& 0.41\\ \hline
Canada& 73.6& 292.6& 0.25\\ \hline
Netherlands& 102.4& 276.4& 0.37\\ \hline
\end{tabular}
\label{tab:country_pub}
\end{table}

\subsection{Topic modeling and clustering}

The dynamic topic model (DTM) conceptualizes the knowledge in nuclear fusion research~\cite{blei2006dynamic,gerrish2010language}. The DTM specifies topics in a set of documents based on latent Dirichlet allocation (LDA)~\cite{blei2003latent}, and it also describes the temporal evolution of detected topics by updating consequent input hyperparameters $\alpha_{t}$ and $\beta_{t}$ by each year. $\alpha_{t}$ affects the topic distribution of a document, and $\beta_{t}$ indicates the word distribution in a topic. The DTM infers both parameters to reproduce the empirical word distribution under the assumption that a document is made by both processes in year $t$, choosing a topic for a document by $\alpha_{t}$ and sampling words in that topic by $\beta_{t}$. $\alpha_{t}$ and $\beta_{t}$ are used as references to estimate $\alpha_{t+1}$ and $\beta_{t+1}$.

In our DTM implementation, insignificant words were filtered out if their term frequency–inverse document frequency (tf-idf) values were less than 0.01. Then, we used the words that appeared more than 10 times in the whole document. As a result, our dictionary contained 7,851 unique words, and the documents contained 1,619,233 words in total. The number of topics $K$ needed to be determined before running the DTM. Following the recent approach \cite{gerow2018measuring}, we specified the number of topics $K = 41$ (see S1 Appendix). Open source codes were written by the authors of the DTM paper and available at \href{https://github.com/blei-lab/dtm}{https://github.com/blei-lab/dtm}. We manually labelled 41 topics from their word frequencies (see S2 Table).

The DTM provides an article's topic distribution based on the learned parameters. As we set the number of topics to 41, the topic distribution of an article was given as a vector of length 41. Topic distribution was allocated to countries in proportion to their contributions on each article. For instance, if an article was written by American authors only, the topic distribution of the article was fully given to the United States. For another article written by three American and two Korean researchers, 60\% of the topic distribution would be added to the United States. In this way, a country's research capability over 41 topics was estimated for each year from 1976-2016.

\subsection{Fractional publication and collaboration counts by topics}

\begin{figure}[!ht]
	\centering
	\includegraphics[width=\textwidth]{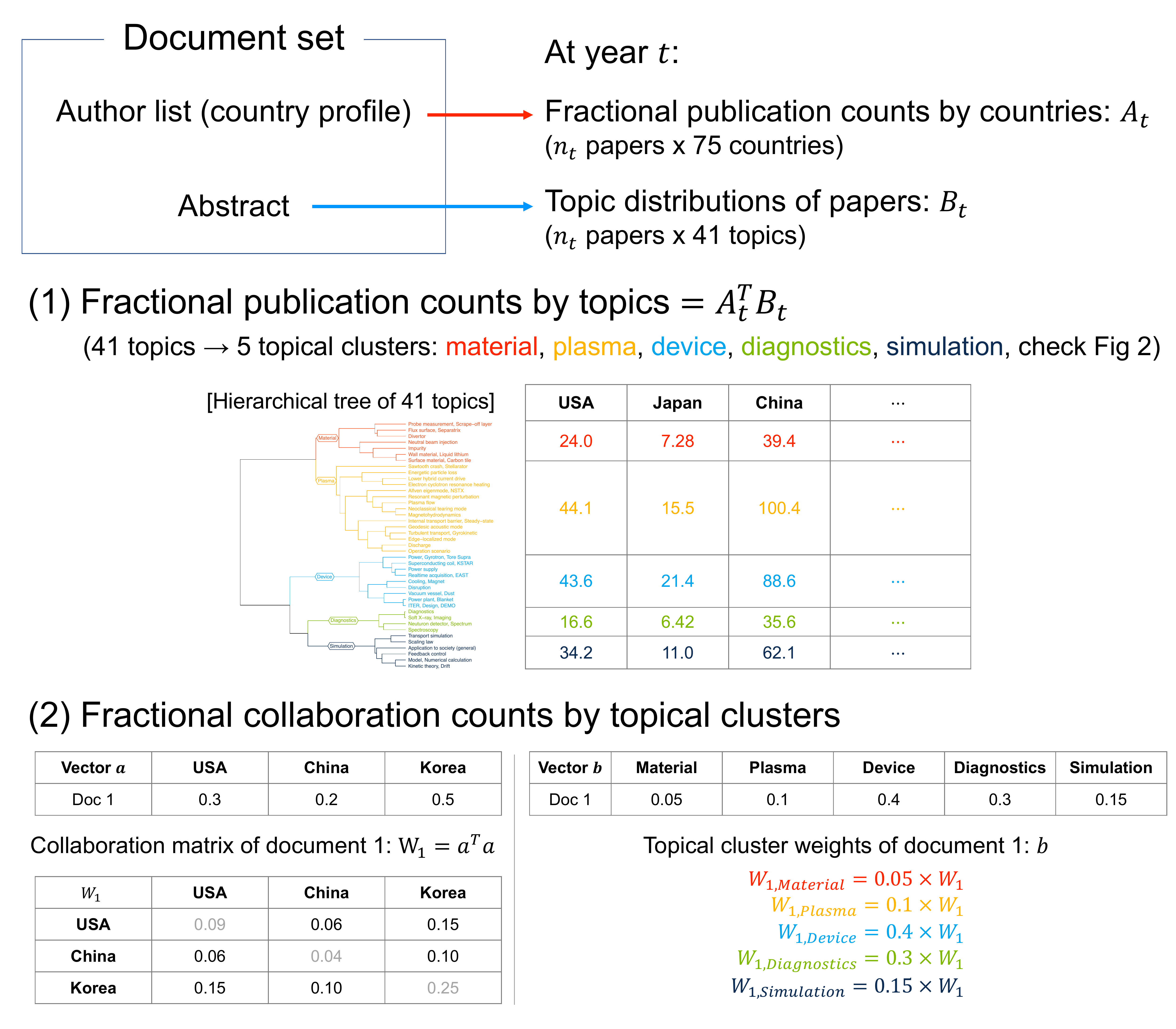}
	\caption{Schematics of the fractional counting method for publication and collaboration counts. Two matrices, the fractional publication counts by countries $A_{t}$ and the topic distributions of papers $B_{t}$, were extracted from the document set of year $t$. (1) $A_{t}^{T}B_{t}$ represents the fractional publication counts by topics at year $t$. For further analysis, based on the hierarchical tree of clusters in Fig \ref{fig:topic_clustering}, the fractional publications by 41 topics are grouped into five topical clusters: material, plasma, device, diagnostics, and simulation. $R_{t}$ is the aggregated matrix and is transposed in the figure to match with the hierarchical tree of 41 clusters. (2) The country profile of a paper is transformed into a collaboration matrix $W_1$, which was distributed over the five topical clusters by weights. For each year, by aggregating the collaboration matrices of all published papers, we had five fractional collaboration matrices.}
	\label{fig:fractional_schematic}
\end{figure}

The fractional counting method was used for calculating a country's publication and collaboration counts (Fig \ref{fig:fractional_schematic}). For year $t$ when $n_{t}$ papers are published, we have two matrices, the fractional publication counts by countries ($A_{t}$: $n_{t}$ papers $\times$ 75 countries) and the topic distributions of papers ($B_{t}$: $n_{t}$ papers $\times$ 41 topics). $A_{t}^{T}B_{t}$ represents the fractional publication counts of 75 countries by 41 topics at year $t$. Based on the five topical clusters that we found (Fig \ref{fig:topic_clustering}), the fractional counts were summed into five columns to obtain the discriminant power for further analysis. We will explain these topical clusters in the result section. We hereafter call this summarized matrix as national research capability over 5 topical clusters at year $t$, $R_{t}$ (75 countries $\times$ 5 topical clusters). Collaborations were also counted in fractions. We multiplied the country profile of a paper and its transpose to obtain the collaboration matrix. The matrix was distributed over five matrices in proportion to topical cluster weights.

\subsection{Normalized revealed comparative advantage (NRCA)}

Normalized revealed comparative advantage (NRCA) \cite{yu2009normalized}, one of revealed comparative advantage indices, represents how much an entity's value exceeds expectations. When comparing longitudinal RCA values, NRCA outperforms the Balassa index (BRCA) \cite{balassa1965trade}, the most popular RCA index that defines comparative advantage as a ratio of observations to expectations. Let $R^{i}_{j,t}$ be country $i$'s research capability on topical cluster $j$ at year $t$. $NRCA^{i}_{j,t}$, the NRCA of country $i$ on topical cluster $j$ at year $t$, is calculated as

\begin{equation}
\label{eq1}
    NRCA^{i}_{j,t} = \Delta R^{i}_{j,t}/R_{t} = (R^{i}_{j,t}-R^{i}_{t}R_{j,t}/R_{t})/R_{t} = R^{i}_{j,t}/R_{t}-R^{i}_{t}R_{j,t}/R_{t}^{2},
\end{equation}

where $R^{i}_{t}$ is the sum of country $i$'s research capability across five topical clusters at year $t$ ($R^{i}_{t}=\sum_{j} R^{i}_{j,t}$), $R_{j,t}$ is the sum of all countries' research capabilities on topical cluster $j$ at year $t$ ($R_{j,t}=\sum_{i} R^{i}_{j,t}$), and $R_{t}$ is the sum of all countries' research capabilities on five topical clusters at year $t$, denoted by $R_{t}=\sum_{i,j} R^{i}_{j,t}$. A positive $NRCA^{i}_{j,t}$ value means that country $i$ has a comparative advantage on topical cluster $j$ at year $t$. 

Countries have comparative advantages on different topics as it is almost impossible to be competitive in all topics. We measured how similar two countries' research capabilities are as follows. First, the NRCA of each country was transformed into the binary vector $\overline{NRCA}$ by changing positive NRCA values to 1 and negative values to 0 to identify the topics with significant comparative advantages. Second, the Jaccard distance between two countries' binary NRCA vectors was calculated for determining their topical dissimilarity (Eq~\ref{eq2}). We call this distance between country $m$ and $n$ on topical cluster $j$ at year $t$ the capability distance $c_{mn,j,t}$. A high $c_{mn,j,t}$ represents that two countries are in complementary relation where their differences in research capability generates synergy by collaborations.

\begin{equation}
\label{eq2}
    c_{mn,j,t} = 1 - \frac{|\overline{NRCA}^{m}_{j,t}\cap\overline{NRCA}^{n}_{j,t}|}{|\overline{NRCA}^{m}_{j,t}\cup\overline{NRCA}^{n}_{j,t}|}
\end{equation}

\subsection{Gravity model of scientific collaboration}

Scientific collaboration between country $m$ and $n$ in topical cluster $j$ at year $t$, $w_{mn,j,t}$, is related to the number of publications of the two ($P_{m,j,t}$ and $P_{n,j,t}$) and their geographical distance ($d_{mn}$). The gravity model explains their relationships in many scientific fields~\cite{ponds2007geographical,hoekman2010research}. $P_{m,j,t}$ and $P_{n,j,t}$ positively and $d_{mn}$ negatively affects $w_{mn,j,t}$. We added the capability distance to the gravity model for checking whether complementarity increases collaboration. Our basic model is written as

\begin{equation}
\label{eq3}
ln(w_{mn,j,t}) \sim \alpha ln(P_{m,j,t})+\beta ln(P_{n,j,t})+\gamma ln(d_{mn})+\lambda c_{mn,j,t},
\end{equation}

where $d_{mn}$ is the Haversine great circle distance (km) between capitals. 
For two countries $m$ and $n$, we counted $w_{mn,j,t}$, $P_{m,j,t}$, and $P_{n,j,t}$ in real values, and calculated $c_{mn,j,t}$ from the binary transformed NRCA vectors. A positive $\lambda$ indicates that complementarity stimulates collaboration.

\section{Results}

\subsection{Knowledge structure of nuclear fusion research}

\begin{figure}[!ht]
	\centering
	\includegraphics[width=0.9\textwidth]{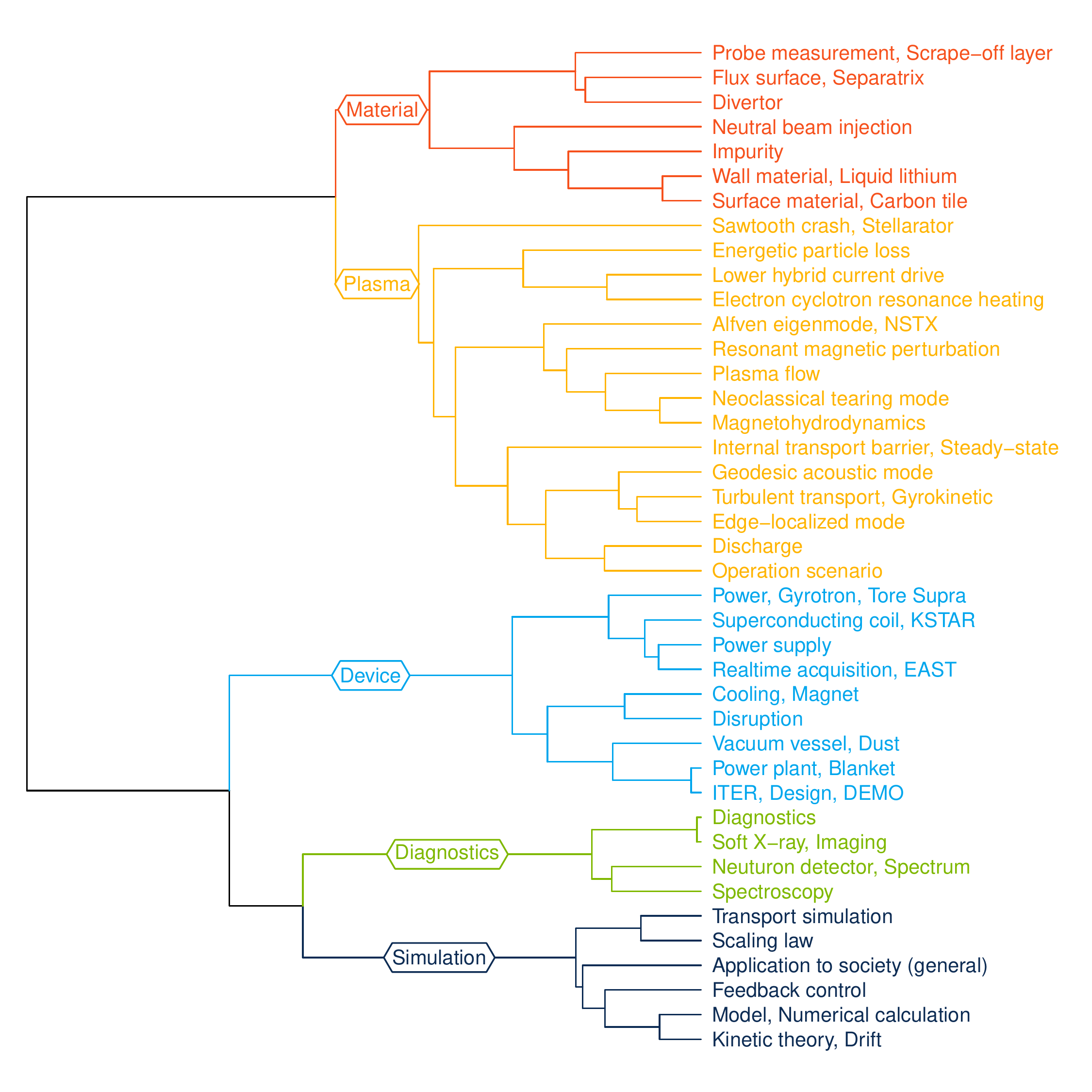}
	\caption{Hierarchical tree of 41 topics detected from the dynamic topic model. Topics were agglomerated by the ward.D method \cite{murtagh2014ward}. The distance between topics was measured by the Jensen-Shannon distance \cite{endres2003new}, a square root of the Jensen-Shannon divergence. Five topical clusters -- material, plasma, device, diagnostics, and simulation -- are revealed. The branches are colored by the corresponding topical clusters.}
	\label{fig:topic_clustering}
\end{figure}

The DTM detected 41 topics in the dataset. Each topic had its word distribution indicating the extent of word assignments to the topic. We assumed that two topics were close if their word distributions were similar. The topic distance between topic $k_1$ and $k_2$ was obtained by the Jensen-Shannon distance~\cite{endres2003new}, a square root of the Jensen-Shannon divergence. For simplicity, we used the word distribution at the last year, $\beta_{2016,k_1}$ and $\beta_{2016,k_2}$. A knowledge structure of nuclear fusion research was drawn by agglomerating 41 topics with the ward.D method~\cite{murtagh2014ward}. The hierarchical tree consists of five distinguishable topical clusters: material, plasma, device, diagnostics, and simulation (Fig \ref{fig:topic_clustering}).

Each cluster is clearly characterized by its topics. We observe the details of each branch from the top of the tree. The ``material'' cluster is described by tokamak edge plasmas and components as plasmas interact with wall materials at the edge. The ``plasma'' cluster contains general plasma-related topics (i.e., plasma flow, magnetohydrodynamics, and discharge), major instabilities in tokamak configurations (i.e., Alfv\'en eigenmode, neoclassical tearing mode, and edge-localized mode), and heating methods (i.e., lower hybrid current drive and electron cyclotron resonance heating). The ``device'' cluster includes mechanical components in tokamaks (i.e., coil, power supply, vessel, magnet, and blanket) and several tokamaks (i.e., Tore Supra, KSTAR, and EAST). The ``diagnostics'' cluster is composed of plasma diagnostics methods such as soft X-ray, neutron detector, and spectroscopy. Finally, the ``simulation'' cluster focuses on analytic calculations and computations.

\subsection{National research capability and its overall trends}

\begin{figure}[!ht]
	\centering
	\includegraphics[width=0.8\textwidth]{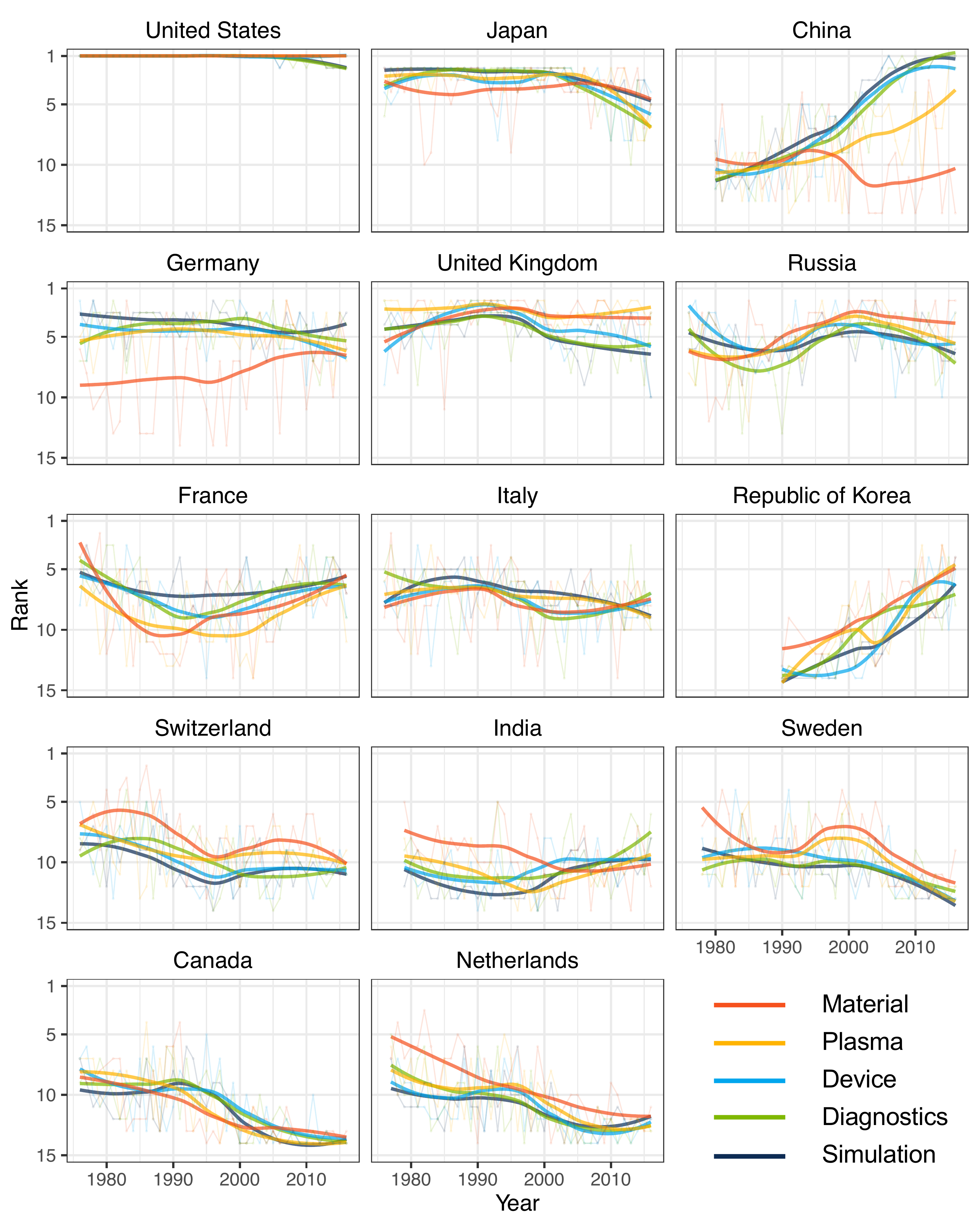}
	\caption{Ranks of normalized revealed comparative advantages for the top 14 countries. Rank series of the countries are smoothed with LOESS (locally estimated scatterplot smoothing) and colored by the topical clusters.} 
	\label{fig:NRCA_rank}
\end{figure}

Normalized revealed comparative advantage (NRCA) on the fractional publication counts extracted national research capability over 40 years (Fig \ref{fig:NRCA_rank}). In all countries, NRCA changes are in good agreement with tokamak construction and operation, representing the scientific effects of large facilities across multiple domains. The United States and Japan have led nuclear fusion research, while Japan's influence has been decreasing since the 2000s. It may be due to the upgrade of their major tokamak JT-60 which was disassembled in 2009-2012 and is being upgraded to JT-60SA for first plasma in 2020. China rapidly develops research capability overall except in material-related topics. Even though we consider the rise of China in all science and technology fields, their pace in nuclear fusion research is surprisingly fast. China's tokamaks, HT-7 and HL-2A, raise research capability in device, diagnostics, and simulation. At the point of EAST (Experimental Advanced Superconducting Tokamak) operation in 2006, they also began to equip plasma capability as well. The other countries operating their own tokamaks, Germany, the United Kingdom, Russia, France, Italy, and Switzerland, actively engage in nuclear fusion research. However, the countries without their own tokamak operation, Sweden and the Netherlands, are losing their research capabilities. Canada's fall seems plausible as they left tokamak projects in the early 2000s~\cite{brumfiel2003canada}. There are two interesting countries, the Republic of Korea and India, that obtain research capability in all fields. Their rises coincide with the ITER project and construction of tokamaks, KSTAR (first plasma in 2008) and SST-1 (first plasma in 2013). 

\subsection{Negative relation between complementarity and collaboration}

Complementarity positively affects collaboration in many science fields~\cite{oh2005coauthorship,barjak2008international,heinze2008across,acosta2011factors,zhang2017china}. Researchers and countries find collaborators that exchange knowledge as well as resources they do not have. We assume complementarity boosts collaboration even in big science because countries have limited budgets and manpower. To observe whether our assumption holds, we implemented the gravity model of collaboration with the capability distance, a Jaccard distance of the binary NRCA vectors in five topical clusters (Eq~\ref{eq3}). The OLS regression results with fixed time effects are given in Table \ref{tab:gravity}. The coefficients of publication counts of two countries are the same because they are symmetric in the collaboration matrix. 

\begin{table}[!ht] 
\caption{Gravity model OLS regression results.\\}
\footnotesize
\begin{threeparttable}
\begin{tabular}{|p{2cm}|p{2cm}|p{2cm}|p{2cm}|p{2cm}|p{2cm}|}
\hline
\textbf{Variables} & \textbf{Material} & \textbf{Plasma} & \textbf{Device} & \textbf{Diagnostics} & \textbf{Simulation} \\ \hline
$ln(P_{m,j})$& 0.497***\newline(0.033)&0.508***\newline(0.032)&0.411***\newline(0.030)&0.438***\newline(0.033)&0.488***\newline(0.033) \\ \hline
$ln(P_{n,j})$& 0.497***\newline(0.033)&0.508***\newline(0.032)&0.411***\newline(0.030)&0.438***\newline(0.033)&0.488***\newline(0.033)\\ \hline
$ln(d_{mn})$& -0.495***\newline(0.044)&-0.451***\newline(0.040)&-0.464***\newline(0.042)&-0.546***\newline(0.049)&-0.485***\newline(0.043)\\ \hline
$c_{mn,j}$& -0.133\newline(0.222)&-0.911***\newline(0.284)&-0.949***\newline(0.232)&-0.690***\newline(0.175)&-0.027\newline(0.194)\\ \hline
Observations&3518&3518&3518&3518&3518\\ \hline
$R^2$&0.113&0.123&0.101&0.094&0.107\\ \hline
\end{tabular}
\begin{tablenotes}[para,flushleft]
     \raggedright Standard error is in parenthesis. Fixed time effects are included. \\
     \raggedright * p-value $<$ 0.1, ** p-value $<$ 0.05, *** p-value $<$ 0.01
\end{tablenotes}
\end{threeparttable}
\label{tab:gravity}
\end{table}

In all topical clusters, as expected, the number of publications had a positive coefficient, and the geographical distance had a negative coefficient. This means that collaborations occur frequently when two countries have high research capability and locate closely. In contrast to our assumption, the capability distance negatively affects collaboration, indicating that countries collaborate less if they have research capabilities in different topics. This tendency is found in three clusters, plasma, device, and diagnostics, with respect to fusion reaction in tokamak facilities. Collaborations on material and simulation are not related to the capability distance. The regression results suggest that complementarity would affect collaborations differently by topics in big science. International collaborations in core knowledge fields happen when two countries mutually benefit based on similar research capability. 

\section{Discussion and conclusion}

Large facilities and international collaboration, two core components of big science, were investigated with bibliographic data, the dynamic topic model, and revealed comparative advantage. In this study, we chose nuclear fusion for a case study. Word similarity between topics unfolded the knowledge structure of nuclear fusion comprising five multidisciplinary topical clusters: material, plasma, device, diagnostics, and simulation. Different countries have different comparative advantages over these clusters. The time points that the comparative advantage trend changes match well with tokamak operation. Catching-up countries that have built their own tokamaks have developed their research capability while countries that do not operate a tokamak miss their productivity. 

Revealed comparative advantage can be used as a new indicator of big science project evaluation. Through time series analysis~\cite{granger1969investigating}, we can examine the connections between facility construction and revealed comparative advantages in different topical clusters. The time series analysis addresses whether knowledge spillover occurs in various scales from facilities to countries~\cite{Hameri1997,horlings2012societal,doi:10.1080/10242690600645233}. In addition, with external information such as the amount of funding, the number of employees, and instrument specifications, we can investigate the impact of facility construction and international collaboration in detail. The publishing policy of large facilities also needs to be considered when interpreting the comparative advantage. Large facilities that restrict the publication of academic papers for the purpose of secrecy~\cite{resnik2006openness} have low research capability in our study, relative to others that promote academic publishing. These qualitative factors of facilities require further evaluations to estimate their scientific impacts accurately as the measure for policy making, investment, and education~\cite{hallonsten2013introducing}.

The international collaboration in nuclear fusion was estimated by the gravity model with the capability distance that represents how similar two countries’ research capabilities are. The regression results show high capability distance distracts the international collaborations in fusion reaction related clusters: plasma, device, and diagnostics. This tendency contrasts with that of other science fields favoring collaborators that have complementary comparative advantages~\cite{oh2005coauthorship,barjak2008international,heinze2008across,acosta2011factors,zhang2017china}. Real collaborations in nuclear fusion governed by this pattern are worth studying. Countries may have distinct motivations to collaborate with other countries and to participate in international projects. Political and societal factors would also be involved in the policy making process. Understanding the history of nuclear fusion research gives us insights into what science policy a country has to take depending on the development stage. 

Our approach can be applied to other fields of big science. Particle physics and Antarctic science are the potential targets. They depend on large facilities, particle accelerators, and research stations in Antarctica. In particle physics, we expect that the dynamic topic model differentiates various types of particle accelerators~\cite{wiedemann2015particle}. A country's strategic decisions for particle accelerators can be traced with comparative advantages on topical clusters. In Antarctic science, research stations may increase research capabilities on geography-dependent topics~\cite{fogg1992history,kim2016bibliometric} because its location expands the range of research activities. An increasing comparative advantage on spatial topics will support this idea. Antarctic science, especially, has interesting aspects that affect the gravity model of collaboration. Collaboration in Antarctica would occur frequently between close research stations, not between close capitals, so the geographical distance of the model should be defined in a different way. The Antarctic Treaty System, which enforces the peaceful usage of Antarctica and freedom of scientific investigation~\cite{berkman2011science}, can encourage countries to collaborate with others having complementary comparative advantages. It is necessary to determine in particle physics and Antarctic science whether collaboration in big science decreases by complementarity as in the case of nuclear fusion. More studies are needed to understand the nature of big science.

\section*{Acknowledgments}
This work was supported by Basic Science Research Program through the National Research Foundation of Korea (NRF) funded by the Ministry of Education (2016R1D1A1B03932590). H.K. acknowledges the NRF Grant funded by the Korean Government (NRF-2017H1A2A1044205, Global Ph.D. Fellowship Program).

\bibliography{main}

\pagebreak
\widetext
\textbf{S1 Appendix. Determining the number of topics from static LDA model.\\}

The recent work using the regression-based document influence model (rDIM) introduces a method to determine the number of topics $K$ as an input of topic modeling~[1]. In general, it runs a static LDA for a large $K$, and then it specifies the number of significant topics whose corresponding documents have a sufficient number of words larger than $w_{th}$. In addition to that, we found the minimal $K$ by varying the threshold $w_{th}$. The details are as follows.

First, we ran a static LDA for $K=500$ following the reference. In each topic $t$ with $n_{d}(t)$ corresponding document, we found the number of documents $n_{x}(t)$ that contained more than $w_{th}$ words (tokens). Then, we determined the significance of the topic from the proportion $p(t) = n_{x}(t)/n_{d}(t)$. In the range near the average value of per-document tokens, $p(t)$ has a Gaussian distribution. From the kernel density estimation (KDE) of this Gaussian distribution, we determined the number of significant topics whose proportion of document $p(t)$ is larger than the cutoff proportion, where the derivative of KDE is minimal. By considering the size of tokens in a document, we set the threshold $w_{th} = 50$. As a result, the number of topics was determined as $K=41$ (S1 FIG).

\begin{figure}[ht]
	\includegraphics[width=0.6\textwidth]{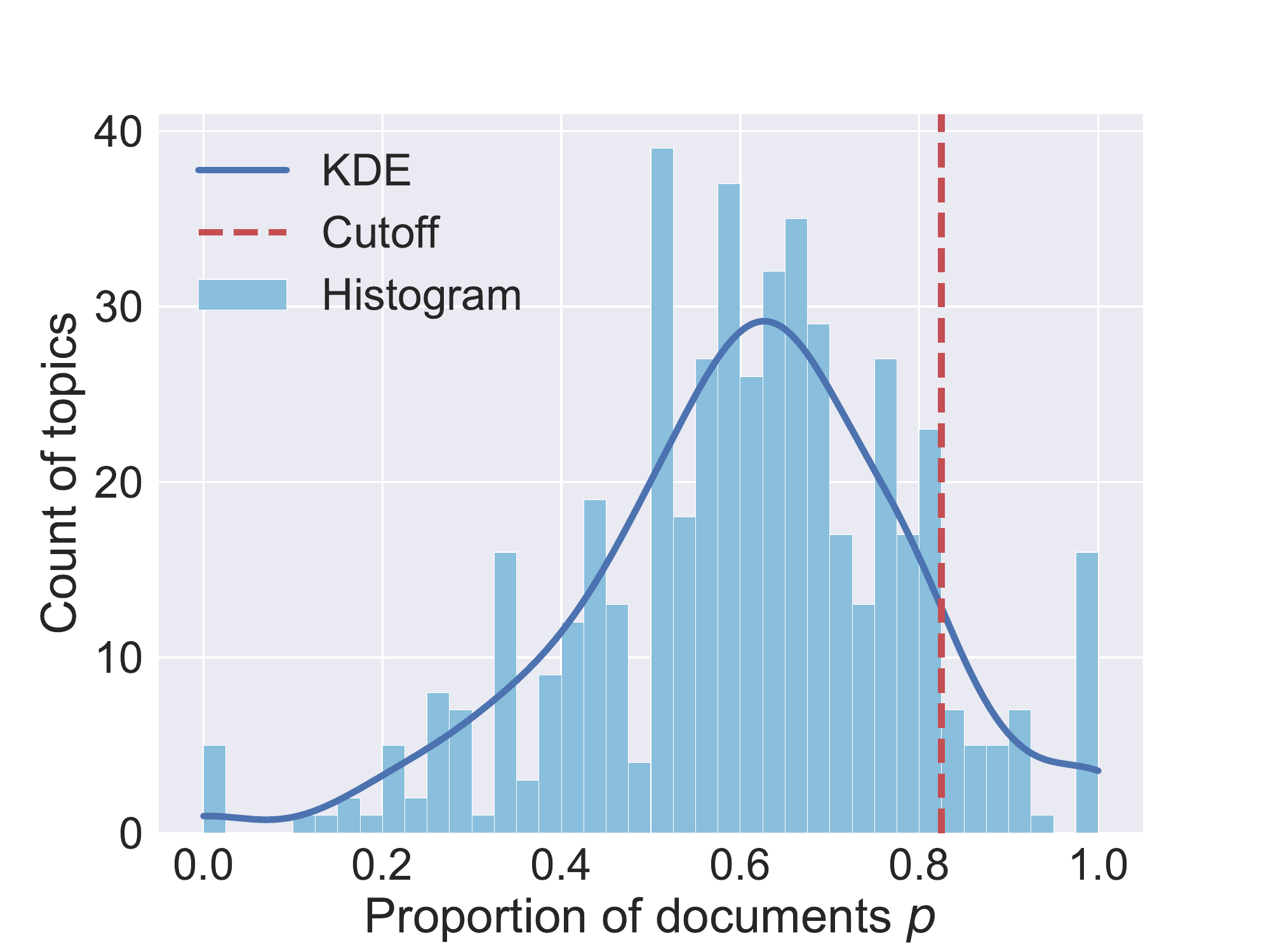}\\
\end{figure}
\noindent S1 FIG: Topic usage distribution for static LDA model. 
We used the topic usage distribution for static $K=500$ model to calculate the cutoff that specifies sufficiently used topics. The minimum of KDE (blue line) derivative determines the cutoff (red dashed line), and the number of topics above this point, $K=41$, is used for the DTM.

\clearpage
\textbf{S2 Table. The top 10 words for 41 topics in nuclear fusion research.\\}

The top 10 words for each topic were obtained from the dynamic topic model on 25,085 abstracts between 1976 and 2016. We manually named 41 topics using the list of the top 15 words of each topic, and the top 10 words among them are listed here.
\begin{table}[h]
\tiny
\centering
\begin{tabular}{|l|l|}
\hline
{\bf Topic} & {\bf Top 10 words} \\
\hline
\hline
Wall material, Liquid lithium            & wall, material, lithium, tungsten, component, heat, surface, pfc, facing, liquid                        \\ \hline
Neuturon detector, Spectrum              & neutron, detector, jet, measurement, reaction, mev, spectrum, rate, fast, radiation                     \\ \hline
Neoclassical tearing mode                & island, tearing, mhd, instability, neoclassical, surface, mn, magnetohydrodynamic, phase, activity      \\ \hline
Transport simulation                     & model, code, transport, agreement, calculation, numerical, developed, simulated, modelling, compared    \\ \hline
Sawtooth crash, Stellarator              & sawtooth, crash, helical, stellarator, reconnection, oscillation, lhd, device, configuration, large     \\ \hline
Neutral beam injection                   & beam, neutral, injection, nbi, source, gas, power, injector, efficiency, injected                       \\ \hline
Resonant magnetic perturbation           & perturbation, resonant, rmp, response, pellet, applied, rmps, coil, toroidal, torque                    \\ \hline
Flux surface, Separatrix                 & surface, side, region, line, poloidal, inside, separatrix, closed, near, midplane                       \\ \hline
Plasma flow                              & flow, toroidal, electric, radial, velocity, poloidal, parallel, zonal, asymmetry, neoclassical          \\ \hline
Surface material, Carbon tile            & surface, tungsten, sample, layer, carbon, deuterium, hydrogen, retention, material, film                \\ \hline
Power, Gyrotron, Tore Supra              & power, antenna, system, ghz, rf, mw, gyrotron, tore, supra, transmission                                \\ \hline
Magnetohydrodynamics                     & stability, equilibrium, pressure, mhd, ideal, profile, beta, ballooning, kink, toroidal                 \\ \hline
Internal transport barrier, Steady-state & state, shear, barrier, transport, steady, profile, formation, itb, reversed, bootstrap                  \\ \hline
Model, Numerical calculation             & method, equation, solution, equilibrium, problem, numerical, function, approach, boundary, distribution \\ \hline
Impurity                                 & impurity, runaway, radiation, discharge, gas, disruption, generation, wall, carbon, injection           \\ \hline
Diagnostics                              & system, measurement, diagnostic, resolution, signal, profile, laser, measure, scattering, spatial       \\ \hline
Divertor                                 & divertor, heat, target, configuration, power, plate, load, outer, particle, lower                       \\ \hline
Power supply                             & power, voltage, supply, system, circuit, loop, arc, breakdown, pulse, kv                                \\ \hline
Discharge                                & increase, time, decrease, increasing, increased, value, discharge, observed, change, rate               \\ \hline
Lower hybrid current drive               & wave, drive, hybrid, lower, lhcd, power, efficiency, frequency, rf, antenna                             \\ \hline
Soft X-ray, Imaging                      & xray, camera, reconstruction, measurement, image, imaging, profile, diagnostics, soft, emission         \\ \hline
Turbulent transport, Gyrokinetic         & transport, profile, gradient, heat, core, particle, turbulent, gyrokinetic, neoclassical, region        \\ \hline
Edge-localized mode                      & elm, hmode, filament, localized, asdex, observed, upgrade, jet, phase, frequency                        \\ \hline
Alfv\'en eigenmode, NSTX                   & spherical, ratio, alfven, aspect, nstx, toroidal, frequency, eigenmodes, torus, gap                    \\ \hline
Electron cyclotron resonance heating     & cyclotron, resonance, ecrh, harmonic, ec, frequency, icrf, power, ech, emission                         \\ \hline
Geodesic acoustic mode                   & fluctuation, frequency, gam, amplitude, geodesic, radial, acoustic, structure, observed, correlation    \\ \hline
Probe measurement, Scrape-off layer      & probe, sol, layer, scrapeoff, measurement, potential, limiter, measured, blob, langmuir                 \\ \hline
Kinetic theory, Drift                    & kinetic, rate, linear, drift, growth, gyrokinetic, instability, gradient, regime, model                 \\ \hline
Spectroscopy                             & line, emission, spectrum, intensity, spectral, charge, spectroscopy, measurement, nm, measured          \\ \hline
Realtime acquisition, EAST               & system, data, realtime, east, acquisition, developed, software, signal, operation, time                 \\ \hline
Application to society (general)         & device, role, physic, interaction, process, play, discussed, particular, understanding, application     \\ \hline
Feedback control                         & wall, feedback, coil, system, controller, position, model, vertical, shape, algorithm                   \\ \hline
Scaling law                              & parameter, scaling, width, power, value, factor, data, law, model, database                             \\ \hline
Vacuum vessel, Dust                      & vacuum, vessel, dust, system, tritium, iter, hydrogen, gas, chamber, safety                             \\ \hline
Energetic particle loss                  & particle, loss, fast, energetic, orbit, distribution, fastion, dust, alpha, ripple                      \\ \hline
Cooling, Magnet                          & cooling, heat, conductor, magnet, strand, helium, superconducting, test, flow, cable                    \\ \hline
Disruption                               & disruption, thermal, force, load, iter, analysis, stress, structure, reactor, method                    \\ \hline
Power plant, Blanket                     & reactor, blanket, design, power, system, analysis, tritium, neutron, module, nuclear                    \\ \hline
ITER, Design, DEMO                       & iter, design, development, system, reactor, demo, physic, component, project, device                    \\ \hline
Operation scenario                       & scenario, operation, power, discharge, drive, performance, iter, profile, limit, regime                 \\ \hline
Superconducting coil, KSTAR              & coil, superconducting, tf, kstar, magnet, system, design, toroidal, vacuum, pf                          \\ \hline
\end{tabular}
\end{table}

\end{document}